# A closer look at low mass post-AGB late thermal pulses

T.M. Lawlor
Pennsylvania State University, Brandywine, Department of Physics, Media, PA 19063 USA

**ABSTRACT**

Late Thermal Pulse (LTP) stellar evolution models experience a helium pulse that occurs following Asymptotic Giant Branch (AGB) departure and causes a rapid looping evolution in the HR Diagram between the Asymptotic Giant Branch (AGB) and Planetary Nebula phase (PN). The transient LTP phases only last decades to centuries while increasing and decreasing in temperature, luminosity, and size over orders of magnitude. LTP objects have often been described in the context of their more dramatic counterparts, very late thermal pulses (VLTP). LTP stars do not evolve as quickly and do not become as hydrogen deficient as VLTP objects. They do not become conspicuous until after resembling a Planetary Nebula for thousands of years. We present stellar evolution calculations from the AGB to the PN phase for models over a range of metallicities from, $Z = 0.0015$ through $Z = 0.03$, and for masses 0.90 $M_\odot$, 1.2 $M_\odot$, and 2.0 $M_\odot$. We focus in on our most dense series (1.2 $M_\odot$, $Z = 0.015$) and designate a stratification of late thermal pulse types based on at what temperature they erupt, which may hint at the progenitor mass. We discuss one type that fits neither a LTP nor VLTP, which may offer an explanation for the star FG Sge. We present the timescales during which LTP models heat up until they reach peak helium burning luminosity, during the rapid luminosity decline, and during the period of cooling and brightening, and we briefly discuss four LTP candidates.

## 1. INTRODUCTION

Few stages of stellar evolution that can be observed in real time, so in general we rely on modelling computations and connecting phases of evolution between different observed stellar phases. Late Thermal Pulse (LTP) stars are one of the few that can in principle be observed in real-time. These are stars that experience a late helium pulse sometime following Asymptotic Giant Branch (AGB) departure and before reaching the white dwarf cooling track. They are a cousin to the more dramatic and often discussed Very Late Thermal Pulse (VLTP) stars, which most notably include Sakurai's Object, V605 Aql, and perhaps HD 167362 the central star of planetary nebula (CSPN) SwSt 1 (Duerbeck et al., 1997; Asplund et al., 1999; Clayton & de Marco, 1997; Hajduk et al., 2020) They have also been linked to several post-AGB helium deficient objects. VLTP stars are a result of a helium shell eruption that occurs when the star has already reached the white dwarf (WD) cooling track, increasing in drastically in brightness in only weeks. LTP stars have proven difficult to observe and few LTP stars have been definitively identified. One likely case is V839 Ara, the central star of the Stingray Nebula (Peña et al., 2022, Balick et al., 2021, Lawlor, 2021, Schaefer et al., 2021, Reindl et al., 2017, Reindl et al.2014). This is the first star to be observed during the most rapid phase of LTP evolution, marked by a steep decline in luminosity over about 25 years. Following a minimum in luminosity, Reindl et al. (2017) announced that the central star had begun to cool and brighten. There is now also little debate left over the nature of FG Sge being a LTP, having evolved back to the AGB (see Jeffery & Schönberner, 2006). The LTP resides in the same place on the HR diagram as PNe evolving from the AGB toward high temperatures at constant luminosity. Besides the rapid phase of evolution there are also slower evolving phases that last hundreds to thousands of years, complicating their identification. They have been invoked in the context of leading to hydrogen deficient [WC] stars and so-called weak emission line stars (wels). The former is a low mass central star that shares spectral features with massive Wolf-Rayet stars (see Werner & Herwig, 2006 and references therein), and wels are likely a misclassification, grouping together disparate stars (Wiedmann et al., 2021, 2015). The degree of hydrogen deficiency in LTP stars is uncertain.

  LTP evolution has been modelled, usually alongside very late thermal pulse (VLTP) objects. This may be because there are fewer confirmed observations to apply LTP models to. But, LTP stars have been studied through modelling over decades, including the central star and its impact on the nebulae. Paczynski (1971) reported that if the mass of a hydrogen envelope is small, the effective temperature can vary drastically during a helium flash, enough to ionize the nebula. Tylenda (1979) modelled nebulae including post-AGB helium flash effects (not yet referred to as late thermal pulses) and reported that time-dependent effects are expected to occur in ionization bounded nebulae of moderate and high excitation during flashes. In the same paper Tylenda suggested that helium flashes can ionize the nebula and showed that a helium flash can account for O I in the inner shell and O II and N II in the outer shell of a nebula.

  Later, Blöcker and Schönberner (1997) divided the types of helium flashes into three types: A final AGB thermal pulse (AFTP) with a hydrogen envelope mass around 0.01 M☉; a LTP, which occurs while evolving at constant luminosity from the AGB to the PN phase with an envelope hydrogen mass $\approx 10^{-4}$ M☉; and a VLTP which occurs after the model star has entered the white dwarf cooling track. Blöcker (2001) provides a review of such flashes in the context of attempting to explain the origin of Wolf-Rayet central stars and PG1159 stars, which has been suggested are descendants of VLTP and LTP stars (Blöcker 2001; Lawlor & MacDonald, 2001; Miller Bertolami & Althaus, 2006; Rühling et al., 2010; Weidmann, et al., 2020). Herwig (2001) and Werner & Herwig (2006) both presented sample models and a review of both LTP and VLTP stars. Lawlor & MacDonald (2006) divided Blöcker's classifications further depending on the metallicity of the star and whether the model departs the AGB burning hydrogen or helium. There, LTP models are identified as a type III departure, while VLTP models are identified as type II. Lawlor (2021) presents select LTP models specifically in the context of V839 Ara, the CSPN in the Stingray Nebula.

  In section two of this paper, we present five new evolution models and briefly describe their evolutions from pre-main sequence through to the thermally pulsing asymptotic giant branch (TPAGB). In section three we use the five models to evolve more dense grids of post-AGB, including LTP flashes. We use designations for our models characterized by the temperature at which models reach peak helium burning luminosity (LTP peak) and examine trends. We present timescales for key periods in LTP evolution, including from nebulae ionization to maximum temperature, to minimum luminosity, and timescales for the evolution back to the AGB. In this section we also discuss surface compositions for LTP models. In section 4 we provide a brief discussion about possible LTP candidates and in section 5 we summarize our conclusions results.

## 2. EVOLUTION CODE AND STARTING MODELS

We present a new set of evolution models using the stellar evolution code BRAHMA, which is a distant descendent of the code originally written by Eggleton (1971, 1972). Our version of the code has been described in Lawlor et al. (2015, 2008), Lawlor & MacDonald (2006) and in MacDonald et al. (2013), thus we describe it only briefly here. Our models are evolved from the pre-main sequence through the WD cooling track by a relaxation method without the use of separate envelope calculations. It uses a time-dependent adaptive mesh technique similar to that of Winkler, Norman & Newman (1984) in which advection terms are approximated by using second-order upwind finite differences. We use standard mixing length theory (Bohm-Vitense, 1958) with a modification to include radiative losses from convective elements when they are optically thin (Mihalas 1978). Convective mixing is modelled as a diffusion process, with the diffusion coefficient determined from the mixing length theory. Mixing due to convective overshooting is not included. We use OPAL radiative opacities (Iglesias & Rogers 1996) for temperatures (in K) above $\log_{10} T = 3.84$. For temperatures below $\log_{10} T = 3.78$, we use opacities from Ferguson et al. (2005). Mass-loss for cool stars is calculated using a Reimers (1975) formula appropriate to red giant branch stars and in addition we fit to the mass-loss rates of Mira variables and OH/IR sources appropriate to stars that have experienced dredge-up of heavy elements, described in detail in Lawlor & MacDonald (2006). For mass loss during hot star evolution, we use an augmented version of the Abbott formulation (Abbott, 1980, 1982)

Here, we briefly describe properties of our cradle to grave models for pre-main sequence through the core helium flash and through the thermally pulsing AGB (if pulses occur). Our starting models include a span of initial metallicities including Z = 0.0015, 0.015, 0.02 and 0.03. In the first case we use initial masses of 0.90 M$_\odot$, and 1.2 M$_\odot$. In the second and third cases we use a mass of 1.2 M$_\odot$ and for the Z = 0.030 case we use a mass of 2.0 M$_\odot$. We place a focus on and create the densest grid of models for our 1.2 M$_\odot$, Z = 0.015 series. For comparisons, we choose two models that overlap in mass (1.2 M$_\odot$) but have lower Z = 0.0015 and higher Z = 0.020, and we compare a lower mass, lower metal model, and higher mass, higher metal model. We also include two series for our primary model for which we adjust mass loss at two different times at and near the final thermal pulse.

In table 1 we show six timescales for different stages of evolution for each model. The time for the pre-main sequence, $\tau_{PMS}$, is taken as the time from t = 0 until the model reaches a minimum luminosity at the zero-age main sequence. The main sequence time scale $\tau_{MS}$ is the time until the central hydrogen composition is equal to zero. The time from the end of the main sequence to central helium ignition is given as $\tau_{CHeI}$. We define this to be when helium burning luminosity is equal 10% of hydrogen shell burning luminosity. The interval for the core helium flash is $\tau_{CHeF}$. This interval includes the time from helium ignition to the point at which star settles into a thermal equilibrium state following the core flash. We show this interval in figure 1. In the top panel of figure 1 we show the evolution between the helium core flash and the thermally pulsing AGB, while in the bottom panel we expand just the thermally pulsing AGB phases. Also noted in figure 1 are the timescales $\tau_{CHeB}$ and $\tau_{TAGB}$. The first of these is the period of relatively steady core helium burning, which we take to last until the central helium abundance drops to $Y_c = 10^{-5}$ and $\tau_{TAGB}$ is the time from the first helium shell thermal pulse till the end of the AGB. We define the end of AGB to be at a time when the radius drops below approximately the same radius as it had at the end of core helium burning. That is, at the point of arrival on the AGB.

**Table 1** Lifetimes of selected evolution phases described in section 2.

| $M/M_\odot$ | Z | $\tau_{PMS}$ (yr) | $\tau_{MS}$ (yr) | $\tau_{CHeI}$ (yr) | $\tau_{CHeF}$ (yr) | $\tau_{CHeB}$ (yr) | $\tau_{TAGB}$ (yr) |
|---|---|---|---|---|---|---|---|
| 0.90 | 0.0015 | $3.05 \times 10^7$ | $9.54 \times 10^9$ | $7.75 \times 10^8$ | $1.30 \times 10^6$ | $9.55 \times 10^7$ | $8.76 \times 10^5$ |
| 1.20 | 0.0015 | $1.75 \times 10^7$ | $3.30 \times 10^9$ | $4.58 \times 10^8$ | $1.50 \times 10^6$ | $9.54 \times 10^7$ | $1.36 \times 10^6$ |
| 1.20 | 0.015 | $3.00 \times 10^7$ | $6.18 \times 10^9$ | $9.40 \times 10^8$ | $2.67 \times 10^6$ | $1.16 \times 10^8$ | $7.37 \times 10^5$ |
| 1.20 | 0.020 | $3.33 \times 10^7$ | $6.97 \times 10^9$ | $1.27 \times 10^9$ | $1.42 \times 10^6$ | $1.25 \times 10^8$ | $3.38 \times 10^4$ |
| 2.00 | 0.030 | $3.07 \times 10^7$ | $1.37 \times 10^9$ | $1.89 \times 10^8$ | $2.01 \times 10^6$ | $1.27 \times 10^8$ | $7.46 \times 10^5$ |

**Table 2** Number of AGB thermal pulses, average and minimum time between pulses, and time between the AGB and PN phase. Every model except the M = 1.2 $M_\odot$, Z =0.015 and Z = 0.020 models experiences an LTP in the original model without changing the mass loss. All crossing times include an LTP except those two models.

| $M/M_\odot$ | Z | N | $\overline{\Delta t}$ | $\Delta t_{min}$ | $\Delta t_{crossing}$ | $M_{wd}/M_\odot$ |
|---|---|---|---|---|---|---|
| 0.90 | 0.0015 | 4 | $2.45 \times 10^5$ | $1.62 \times 10^5$ | $9.10 \times 10^4$ | 0.548 |
| 1.20 | 0.0015 | 9 | $1.63 \times 10^5$ | $1.09 \times 10^5$ | $2.41 \times 10^4$ | 0.601 |
| 1.20 | 0.0150 | 6 | $1.58 \times 10^5$ | $1.50 \times 10^5$ | $1.40 \times 10^4$ | 0.550 |
| 1.20 | 0.0200 | 1 | — | — | $5.80 \times 10^4$ | 0.546 |
| 2.00 | 0.0300 | 4 | $3.33 \times 10^5$ | $2.00 \times 10^5$ | $1.22 \times 10^4$ | 0.635 |

In table 2 we present the number of pulses, N, for each model during the thermally pulsing AGB. For M = 1.2 $M_\odot$ with metallicity Z = 0.02, the model experiences only one thermal pulse (we do not count the late thermal pulse for this purpose). Higher Z models experience fewer thermal pulses overall (see table 2) and the average time between pulses is shorter for models of the same mass. This is because our models with higher element abundances experience more mass loss and thus fewer pulses, due to metal dependence of our mass loss prescription. Like what was found in Lawlor & MacDonald (2006), the average time between AGB thermal pulses decreases with increasing core mass and for increasing Z. Also in table 2, we present the crossing times for each model. Here, we define crossing time as the time between AGB departure (defined earlier) and maximum temperature. It is important to note here that most of these initial models experience a late thermal pulse by coincidence, which complicates and extends each crossing time by approximately 3000 – 5000 years.

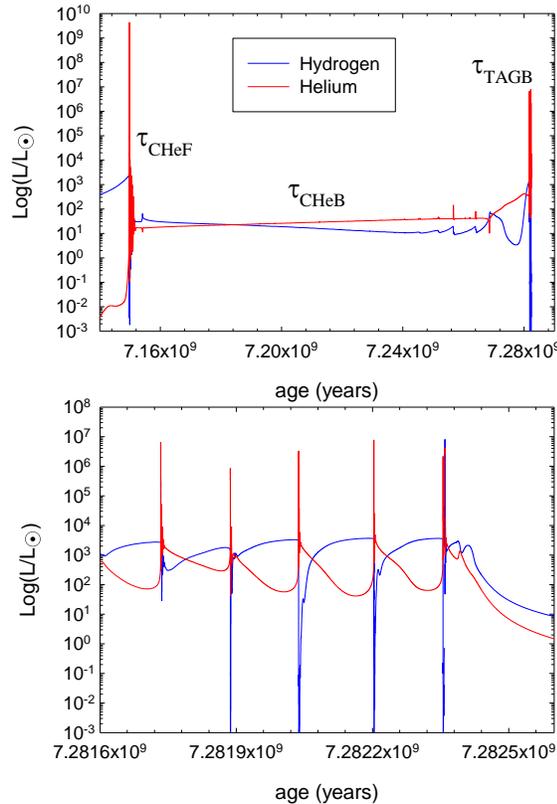

**Figure 1** *Top:* M = 1.2 M$_\odot$, Z = 0.015 hydrogen and helium burning luminosities as a function of age from just before the core helium flash to just after the AGB thermal pulses. *Bottom:* The same as top but for an expanded view of the AGB thermal pulses. We also show one example of a very late thermal pulse, as labeled.

Only our initial M = 1.2 M$_\odot$, Z = 0.015 and Z = 0.020 models evolved to the PN without experiencing an LTP. This is also why our model with M = 1.2 M$_\odot$ and Z = 0.015 does not follow the expected trend of increasing crossing time with increasing metallicity when compared to our model of the same mass and lower Z = 0.0015. We discuss LTP departure timescales in more detail in the following section. Also shown are the final WD masses in each case, and as expected the highest Z models for the same initial mass strips off the most mass by mass loss. In this case for models beginning with M = 1.2 M$_\odot$, the Z = 0.02 model mass is reduced to 0.546 M$_\odot$ as compared to 0.601 M$_\odot$ for our Z = 0.0015 model. In the next section we turn our attention to the crux of this work, post-AGB LTP models.

## 3 POST AGB LATE THERMAL PULSE MODELS

LTP objects are the result of an intense and rapidly evolving helium flash that occurs after the AGB but before entering the WD cooling track. Few stars have been definitively identified as an LTP star. The central star V839 Ara (SAO244567, Hen-57) may be the first LTP to be definitively observed in real time (Reindl et al. 2017; Shaeffer et al. 2021; Lawlor 2021). Indeed, Peña et al. (2022) report that the central star has continued to cool, and is now below 40,000 K. They also note through careful analysis of the surrounding nebula that "the Stingray nebula seems to be a normal non-Type I planetary nebula with sub solar abundances."

### 3.1 Late thermal pulse types

To create a grid of LTP models from each of our five starting models, we adjust model mass loss beginning at the peak of the final helium thermal pulse near the end of the AGB. For our M = 1.2 $M_\odot$, Z = 0.015 model grid, we adjust mass loss at two separate times: one at the peak of the final thermal pulse and one when the helium pulse is declining. The reason for this is that the following behaviour can vary depending on where in the thermal pulse cycle the model star is when it leaves the AGB. In doing so, we create models that cross the HR diagram with a range of masses and particularly a range of helium layer masses, that later experience a LTP. Lawlor & MacDonald (2006) identified a LTP that erupts at temperatures above Log($T_{eff}$) = 4.45 as type III departure. This is approximately the critical temperature above which a star can ionize of the surrounding nebulae. In the current work, all LTP's are type III. We also include what Lawlor and MacDonald classified as type V and define a type II/III for comparison. The former of these models are those that Blöcker (2001) identified as an AGB Final Thermal Pulse (AFTP). Type V or AFTP evolution in the HR-Diagram superficially resembles that of an LTP, but the peak helium burning luminosity for the pulse occurs while still on the AGB and the model star has just begun to contract. The latter, type II/III, experience a helium flash after turning the corner at $T_{max}$ but earlier than typical type II models. A type II departure corresponds to a VLTP. Here, we focus in on type III (LTP) models and make comparisons using the designations type IIIc, IIIb, IIIa, and type II/III, listed in order of increasing temperature range at which a helium flash occurs. Samples of each temperature range designation are presented in an HR-diagram in Figure 2. For type IIIc models, the peak helium burning luminosity occurs when 4.65 < Log($T_{eff}$) < 4.78. The peak for type IIIb LTP models occur when 4.8 < Log($T_{eff}$) < 4.92 and type IIIa LTP model peaks occur when occurs when Log($T_{eff}$) > 4.95. We designate Type II/III to be those that reach peak helium burning luminosity when Log$T_{eff}$ > 5.0, very shortly after turning toward the white dwarf cooling track. As we will discuss, these have similarities to both LTP and VLTP models.

  Although there are some weak parameter associations for these temperature designation types such as mass loss rate and mass of helium, the primary reason we see models flash at the temperatures they do is likely due to the rate of change of temperature as they cross the HR diagram. For example, our 1.2 $M_\odot$ models (with Z = 0.015 and 0.0015) evolve away from the AGB more quickly in temperature (20 - 40 K/yr) until Log$T_{eff}$ = 4.7 and then begin to progressively slow to a rate of 10 - 20 K/yr. Thus, this series is more likely to undergo a helium flash peak after evolving beyond this temperature simply because models spend more time evolving there. Model mass and metallicity can affect at what temperature evolution begins to slow. Models with mass 1.2M$_\odot$ spend about 70 percent of the crossing time above LogT$_{eff}$ = 4.7. Therefore, we see fewer type IIIc flashes than hotter types in those series of models. Those that flash at LogT$_{eff}$ = 4.7 (type IIIc) do not slow down but increase to 80 - 90 K/yr as helium burning increases shortly before the flash, beginning around LogT$_{eff}$ = 4.6. The picture is different for our Z = 0.02, 1.2 M$_\odot$ model. Having a higher Z, that model evolves the slowest (7.0 – 9.0 K/yr) and the rate remains nearly uniform. For our lowest mass 0.9 M$_\odot$ model, the rate of change in temperature begins to slow from 11 K/yr down to 4.0 K/yr at a cooler LogT$_{eff}$ = 4.4, and in this series, we see more of the cooler type IIIc flashes, some type IIIb and no type IIIa. The rate of temperature change for our type II/III models is the most rapid (60 K/yr on average for our 2.0 M$_\odot$ models) until LogT$_{eff}$ = 4.7 and is reduced to 49 K/yr after. Their helium flashes peak later when the rate of temperature change is nearly zero, a few thousand years after turning the corner at T$_{max}$. They spend more than 85 percent of their crossing evolution in this region leading up to the flash. Thus, type II/III flashes are more likely to occur in models that evolve the most quickly across the HR diagram. We note that the temperatures at which an LTP peaks (identifiable by a fast decrease in overall luminosity) may then hint at a star's progenitor mass. A subsequent study of more massive progenitor LTPs may answer this question.

    We show the time evolution of the helium and hydrogen burning luminosity in four panels in figure 3 for all four types. The significance of the peak LTP is that it coincides with a rapid, and observable (20 – 150 years) decline in overall luminosity. Also shown in the first panel in figure 3 and in figure 2 is a sample type V model (AFTP). These models reach peak helium burning luminosity when Log(T$_{eff}$) < 3.71 and temporarily reduce hydrogen burning luminosity before it rebounds and steadily burns until reaching the WD cooling track. Finally shown in the bottom panel of figure 2 is a type II/III model which reaches peak helium burning luminosity immediately following the PN phase as it turns towards the WD cooling track. In type II/III models, there is no hydrogen consumption and in fact hydrogen burning is temporarily extinguished. Type V and II/III models bracket the behaviour of the three types of LTP models.

    In Table 3 we present how remaining helium layer masses change between the end of the AGB to the peak of an LTP (or AFTP and VLTP in the cases of type V and II) for all our post-AGB models. As expected, there is a general correspondence for each individual model series between how much helium remains at AGB departure and what type of LTP results. This is because as the mass of helium increases during the final thermal pulse cycle, it leaves the AGB closer to the next thermal pulse cycle. In general, with some exceptions, lower helium layer masses lead to a type IIIa at one end and higher helium layer masses lead to a type IIIc or V. Having less helium left, type IIIa LTP's occur when the stellar radius is smaller and the overall structure of the star is more dense. We show that in our solar metallicity grid and in our grid with Z = 0.02 of the same mass, that models with a helium layer between 3.3% and 4% of the total stellar mass experience an LTP of some variety. Our single model with less helium, 2.95% helium, reaches the WD cooling track and experiences a VLTP. The range of values for an LTP

depends on the initial mass and metallicity of the models. Our initial model of mass 1.2 $M_\odot$ and Z = 0.0015 and our model with 2.0 $M_\odot$ and Z = 0.03 require helium mass between 1.8% and 2.8% of the total stellar mass to undergo a LTP or type II/III departure in the case of the latter models.

Our 1.2 $M_\odot$, Z = 0.0015 models terminate with a higher mass (~ 0.60 $M_\odot$) than our 1.2 $M_\odot$, Z = 0.015 and our 0.9 $M_\odot$, Z = 0.0015 model. Both have final masses between 0.54 $M_\odot$ and 0.55 $M_\odot$. Our 2.0 $M_\odot$ model also terminates with a higher final mass between 0.62 $M_\odot$ and 0.64 $M_\odot$. In both cases with the highest final masses, LTP helium burning peaks occur more often at higher temperatures. In the first case, four of the seven LTP models are of hotter type IIIa and one is the hottest type II/III. For the latter, 2.0 $M_\odot$ case, three of the five models are the hottest type II/III LTP. For this case, the other two models are type V and type I, and there are no other types in this series. This is again, primarily due to spending less time evolving through cooler temperatures. In contrast, the lowest 0.90 $M_\odot$ model series and the 1.2 $M_\odot$, solar Z models (both with lower final masses between 0.52 $M_\odot$ – 0.54 $M_\odot$) experience a wider distribution of LTP types, and neither exhibit the hottest type II/III. As discussed earlier, this is primarily due to timescales during crossings and hints at higher mass remnants having a predilection toward higher temperature flashes, where the rate of change of temperature is decreased.

For models with helium mass outside of the range of mass that results in a LTP, two different behaviours are observed. For example, models that reach the end of the AGB with less helium than LTP models will either experience a type II/III helium pulse, a VLTP, or no final pulse which evolves directly into the WD cooling track. The first results in a moderately hydrogen deficient WD (see next subsection), a VLTP results in a hydrogen deficient WD, and the later results in a DA WD. At the other end, models that leave the AGB with helium layer mass higher than that for an LTP produces either an AFTP or simply an additional thermal pulse on the AGB, which can then lead to a DA WD. In that case the model evolves to the PN phase and WD cooling track with a marginally lower overall mass. This includes models in the M = 1.2 $M_\odot$, Z = 0.020 series with masses 0.5457 $M_\odot$ and 0.5458 $M_\odot$ (see Table 3). Also shown in Table 3 is mass of hydrogen at the peak of the pulse, which is essentially a proxy for temperature and radius also shown, and we show the ratio of hydrogen to helium at the LTP peak. This ratio is in general lower for models that reach peak helium burning at higher temperatures. In table 4, we summarize ranges of key parameters for each of the types of LTP models.

There are few, but some previously published models that at least qualitatively resemble the types presented here. For example, Blöcker (1995b, fig. 14) presented a 3.0 $M_\odot$ post AGB evolution model which by appearances resembles our type IIIa models and another 5.0 $M_\odot$ initial mass model that results in a flash that resembles our type II/III. Herwig et al. (1999) include a 2.0 $M_\odot$ model that resembles our type II/III models, though there they identify it as a very late thermal pulse (type II). Althaus et al. (2005) invoke a LTP as the reason for some DA white dwarfs having a thin hydrogen envelope, and their 2.7 $M_\odot$ post-AGB model resembles our type V thermal pulse model, though they identify it as an LTP. And, Miller Bertolami (2005) present a similar work as Althaus et al. in which they use a model that more resembles our type IIIa LTP

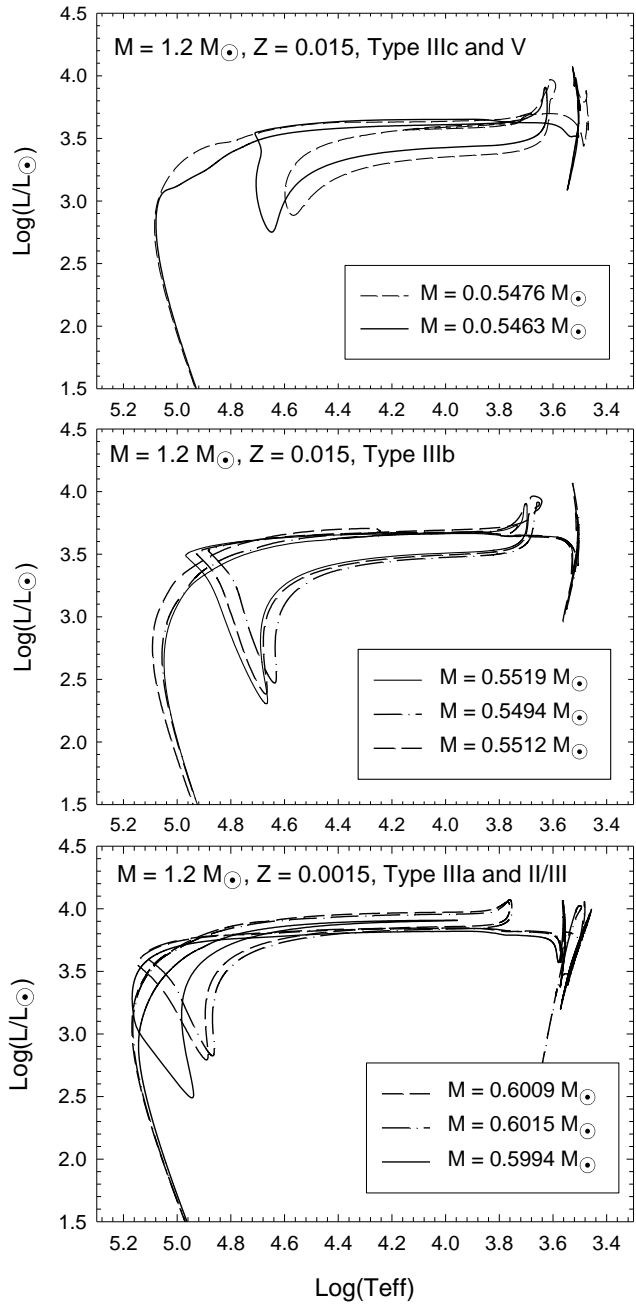

**Figure 2** HR-diagrams for each type of model departure (labeled). In the top panel, the solid line is a type IIIc LTP and the dashed line is type V. In the bottom panel, the solid line is a type II/III and the rest are type IIIa. For LTP models, the peak helium flash (LTP) occurs approximately at maximum temperature before the rapid decline in luminosity. The AFTP peak occurs while the model is still cool, and the type II/III peak occurs after maximum temperature.

**Table 3** Post-AGB stellar model parameters. M/M$_\odot$ post-AGB is the post-AGB stellar model mass, M$_{HE}$/M$_{s\ AGB}$ is the helium layer mass as a fraction of stellar mass taken at AGB departure and M$_{HE}$/M$_{s\ LTP}$ is the same but taken at the LTP peak. M$_H$/M$_{s,\ LTP}$ is the mass of hydrogen as a fraction of stellar mass and M$_H$/M$_{HE,\ LTP}$ is the ratio of hydrogen to helium layer mass, both taken at the LTP peak. Also shown is Log of effective temperature and radius. The column labelled 'Type' is a designator that describes the temperature range in which the LTP peak occurs. For our solar metallicity model, we include two series: one with mass loss adjusted at the final thermal pulse peak, and another with mass loss adjusted during the decline of the final pulse. For type V departure models, the end of the AGB follows the helium pulse (AFTP), thus the mass of helium increases. Values for type V models are taken from the peak of the AFTP rather than LTP. The parameter ε is our mass loss multiplier and may be useful as an identifier.

| Mass | ε | M/M$_\odot$ post-AGB | M$_{HE}$/M$_s$ AGB | M$_{HE}$/M$_s$ LTP | M$_H$/M$_s$ LTP | M$_H$/M$_{HE}$ LTP | Log(T$_{eff}$) LTP | Log(R) LTP (cm) | Type |
|---|---|---|---|---|---|---|---|---|---|
| 0.9 M$_\odot$ | Z = 0.0015 | | | | | | | | |
| | 0.70 | 0.5503 | 0.0405 | 0.0406 | 0.00354 | 0.0872 | 3.63 | 12.99 | V |
| | 1.00 | 0.5512 | 0.0420 | 0.0403 | 0.00134 | 0.0334 | 4.65 | 10.84 | IIIc |
| | 1.20 | 0.5474 | 0.0401 | 0.0387 | 0.00135 | 0.0350 | 4.68 | 10.77 | IIIc |
| | 1.40 | 0.5468 | 0.0394 | 0.0384 | 0.00129 | 0.0337 | 4.72 | 10.67 | IIIc |
| | 1.60 | 0.5465 | 0.0391 | 0.0381 | 0.00118 | 0.0310 | 4.78 | 10.52 | IIIc/b |
| | 0.50 | 0.5513[##] | 0.0325 | 0.0320 | 0.00131 | 0.0411 | 4.82 | 10.43 | V, IIIb |
| | 1.80 | 0.5461 | 0.0385 | 0.0378 | 0.00118 | 0.0313 | 4.81 | 10.47 | IIIb |
| 1.2 M$_\odot$ | Z = 0.0015 | | | | | | | | |
| | 0.50 | 0.6045 | 0.0265 | 0.0249 | 0.00060 | 0.0241 | 4.79 | 10.68 | IIIc/b |
| | 1.30 | 0.6007[#] | 0.0199 | 0.0181 | 0.00055 | 0.0304 | 4.88 | 10.45 | IIIb |
| | 0.70 | 0.6029 | 0.0240 | 0.0233 | 0.00032 | 0.0136 | 5.09 | 9.98 | IIIa |
| | 1.00 | 0.6015 | 0.0215 | 0.0205 | 0.00033 | 0.0162 | 5.09 | 9.96 | IIIa |
| | 1.20 | 0.6009 | 0.0208 | 0.0197 | 0.00032 | 0.0161 | 5.11 | 9.88 | IIIa |
| | 1.50 | 0.6002 | 0.0197 | 0.0188 | 0.00037 | 0.0194 | 5.10 | 9.88 | IIIa |
| | 2.00 | 0.5994 | 0.0186 | 0.0180 | 0.00032 | 0.0176 | 5.10 | 9.68 | II/III |
| 1.2 M$_\odot$ | Z = 0.015 Mass loss after TP peak | | | | | | | | |
| | 0.71 | 0.5417 | 0.0321 | 0.0324 | 0.00189 | 0.0583 | 3.58 | 13.1 | V |
| | 0.70 | 0.5463 | 0.0315 | 0.0305 | 0.00107 | 0.0350 | 4.70 | 10.71 | IIIc |
| | 0.75 | 0.5457 | 0.0320 | 0.0309 | 0.00076 | 0.0247 | 4.87 | 10.32 | IIIb |
| | 0.75 | 0.5460 | 0.0315 | 0.0303 | 0.00076 | 0.0252 | 4.85 | 10.4 | IIIb |
| | 0.90 | 0.5460 | 0.0321 | 0.0310 | 0.00082 | 0.0264 | 4.86 | 10.33 | IIIb |
| | 0.66 | 0.5494 | 0.0343 | 0.0333 | 0.00071 | 0.0212 | 4.87 | 10.38 | IIIb |
| | 0.72 | 0.5512 | 0.0372 | 0.0354 | 0.00067 | 0.0190 | 4.91 | 10.28 | IIIb |
| | 0.60 | 0.5520 | 0.0389 | 0.0374 | 0.00067 | 0.0180 | 4.89 | 10.33 | IIIb |
| | 0.55 | 0.5519 | 0.0393 | 0.0379 | 0.00060 | 0.0158 | 4.95 | 10.19 | IIIa |
| 1.2 M$_\odot$ | Z = 0.015 Mass loss at TP peak | | | | | | | | |
| | 0.40 | 0.5476 | 0.0343 | 0.0347 | 0.00129 | 0.0373 | 3.71 | 12.78 | V |
| | 0.50 | 0.5536 | 0.0376 | 0.0383 | 0.00249 | 0.0652 | 3.64 | 12.92 | V |
| | 0.80 | 0.5469 | 0.0314 | 0.0304 | 0.00105 | 0.0346 | 4.72 | 10.68 | IIIc |
| | 0.50 | 0.5527 | 0.0402 | 0.0389 | 0.00087 | 0.0222 | 4.73 | 10.68 | IIIc |
| | 0.60 | 0.5517 | 0.0363 | 0.0345 | 0.00072 | 0.0209 | 4.85 | 10.41 | IIIb |
| | 0.90 | 0.5504 | 0.0364 | 0.0349 | 0.00055 | 0.0157 | 4.98 | 10.05 | IIIa |
| | 0.55 | 0.5391 | 0.0295 | 0.0277 | 0.00014 | 0.0052 | 4.92 | 9.55 | II |
| 1.2 M$_\odot$ | Z = 0.020 | | | | | | | | |
| | 1.40 | 0.5458 | 0.0328 | — | — | — | — | — | I |
| | 1.20 | 0.5457 | 0.0328 | — | — | — | — | — | I |
| | 0.70 | 0.5460 | 0.0334 | 0.0320 | 0.00098 | 0.0308 | 4.89 | 10.22 | IIIb |
| | 0.90 | 0.5459 | 0.0331 | 0.0321 | 0.00103 | 0.0321 | 4.91 | 10.08 | IIIb |
| | 1.00 | 0.5460 | 0.0330 | 0.0320 | 0.00098 | 0.0307 | 4.92 | 9.95 | IIIb |
| | 0.50 | 0.5463 | 0.0342 | 0.0332 | 0.00061 | 0.0183 | 4.98 | 9.85 | IIIa |
| 2.0 M$_\odot$ | Z = 0.030 | | | | | | | | |
| | 1.00 | 0.6393 | 0.0176 | 0.0174 | 0.00067 | 0.0383 | 3.77 | 12.74 | V |
| | 0.90 | 0.6275 | 0.0187 | 0.0178 | 0.00014 | 0.0080 | 5.01 | 9.49 | II/III |
| | 1.20 | 0.6375 | 0.0180 | 0.0175 | 0.00016 | 0.0092 | 5.07 | 9.51 | II/III |
| | 1.40 | 0.6373 | 0.0178 | 0.0173 | 0.00015 | 0.0089 | 5.10 | 9.52 | II/III |
| | 0.40 | 0.6394[++] | 0.0151 | — | — | — | 3.58 | 13.12 | I |

[++]Mass loss decreased the most for this series, thus it experiences an additional AGB thermal pulse which reduced helium layer mass and then it becomes a DAWD. [#]Model mass loss began after thermal pulse peak. [##]Model experiences an additional helium pulse at Log(T$_{eff}$) = 3.58 on the AGB, then an LTP later.

in appearance. Since these works do not report where the peak helium burning luminosity occurs, we cannot definitively identify them. We caution that other published models described here use larger masses, different initial metallicities, and include effects such as convective overshoot, not included here. However, they do potentially offer some evidence that higher mass models are more likely (not exclusively) to experience a hot LTP.

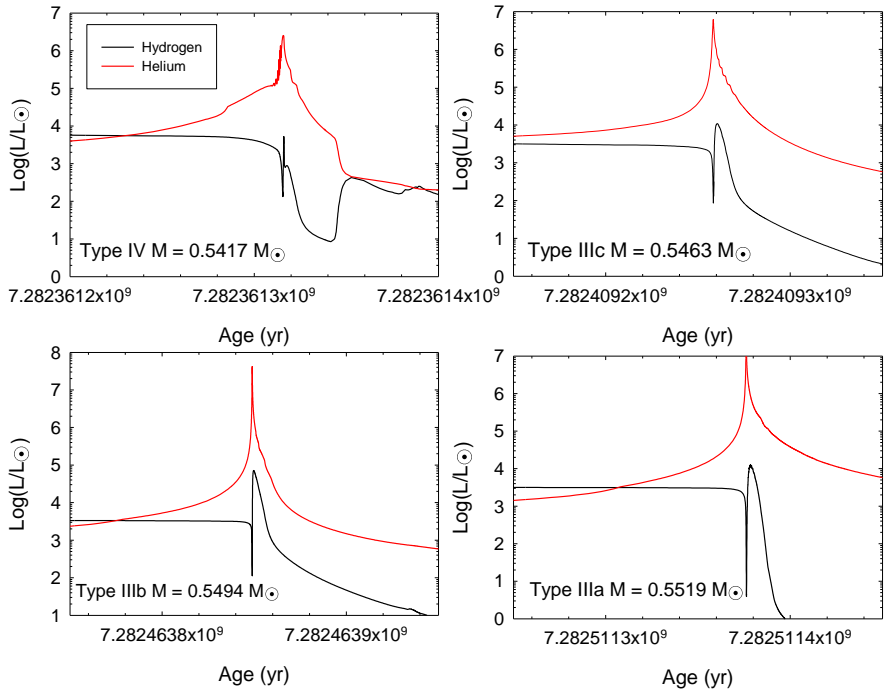

**Figure 3** Hydrogen and helium burning luminosities as a function of age for LTP's of type IIIa through type V. Each horizontal axis spans 200 yr. Secondary hydrogen burning pulse decreases with temperature/type for LTP's.

**Table 4** Summary of the range of parameters for three types of LTP's, type V (AFTP), and type II/III. The ratio of helium to stellar mass is taken at the end of the AGB.

| LTP Type | Log($T_{eff}$)$_{peak}$ | Log(R)(cm)$_{peak}$ | $M_{He}$ / $M_{s, AGB}$ | $M_H$/$M_{He, LTP}$ |
|---|---|---|---|---|
| V (AFTP) | 3.58 – 3.71 | 12.74 – 13.10 | 0.018 – 0.041 | 0.040 – 0.090 |
| IIIc | 4.65 – 4.79 | 10.67 – 10.84 | 0.027 – 0.042 | 0.022 – 0.035 |
| IIIb | 4.80 – 4.92 | 9.95 – 10.47 | 0.019 – 0.039 | 0.019 – 0.041 |
| IIIa | > 4.95 | 9.85 – 10.19 | 0.020 – 0.039 | 0.014 – 0.019 |
| II/III | > 5.00 | 9.49 – 9.86 | 0.018 – 0.019 | 0.008 – 0.018 |

## 3.2 Surface compositions and timescales

Timescales between the end of the AGB and the point when an LTP reaches peak brightness can take 1000s years or more. During much of this time, the central star is hot enough to ionize surrounding material. Thus, identifying LTP stars is problematic, and in fact LTP stars could well be hiding in plain sight resembling PNe for decades or centuries. Detecting LTP's among PNe would require continuously monitoring for declines in brightness or a measurable change in temperature over decades. Otherwise, LTP objects might go undetected until an observer notices they have significantly heated and dimmed or cooled while brightening. This is the period that helped expose V839 Ara as a likely type IIIc LTP (Lawlor, 2021; Reindl et al., 2017; Reindl et al.2014; Schaefer & Edwards, 2015).

      Potentially making detection more problematic, most models do not consume enough hydrogen to become as hydrogen deficient as a VLTP. One of the primary differences for a VLTP is that helium ignites after the star enters the white dwarf cooling track, experiencing significant mixing at the surface, and becoming exceedingly hydrogen deficient. This is because a VLTP star no longer has an entropy barrier during the helium flash, and a majority of the hydrogen layer at the surface is consumed by convection, leaving the star's surface enriched in helium due to dredge up. This change in surface composition is absent or modest in most LTP models presented here. Only a few LTP models that erupt at the highest temperatures and with the smallest radii become modestly enriched in He, C, N and O compared to other LTP model types, but not until later in the evolution. Our type II/III models consume about half the hydrogen. This may be relevant for CSPN that are only modestly H-deficient. This feature may also help identify FG Sge type stars as we discuss in the following section.

      All LTPs presented here experience little surface hydrogen consumption or dredge up, even by the time they reach the WD cooling track. Some models have hydrogen surface abundance reduced from $X = 0.711$ to $0.67$ in mass fractions, and one is reduced to $0.57$ but this occurs prior to the LTP, while on the AGB. The LTP does consume some hydrogen, as is shown in figure 3, but the convection zones do not deplete the stellar surface. Here, we are in agreement with Miller Bertolami (2005), who indicates that there is not significant depletion of hydrogen in their LTP models. However, Blöcker (2001) and Herwig (2001) show helium and carbon enrichment by invoking convective overshoot, and Herwig used a higher mixing length of 3.0. Our mixing length parameter is 1.70. We note that we do not include convective overshoot but it is not clear if overshoot is needed in this context. Continued monitoring of V839 Ara may inform us of this.

      Type II/III thermal pulses are essentially a late LTP (or an early VLTP). It is distinguishable from both types, however. Unlike LTP models, they reach a helium burning luminosity peak very near or just entering the WD cooling track, when the stellar luminosity is already declining, and the radius has become small ($\text{Log}(R(cm)) < 9.80$). In the case of our 2.0 $M_\odot$, type II/III thermal pulses, the surface becomes more hydrogen deficient than all other LTPs. For our 0.6274 $M_\odot$, $Z = 0.030$ model (initial mass 2.0 $M_\odot$) the hydrogen surface abundance is reduced to $X = 0.415$ from $0.711$, and helium is enriched to $Y = 0.552$. This begins to occur when the model track circles back to the AGB and shortly after. Hydrogen is reduced less so in other type II/III models. For example, our 0.6373 $M_\odot$ and 0.6375 $M_\odot$ models reduce hydrogen to $X = 0.60$ and increases helium to $Y = 0.37$. Type II/III pulses are also unlike VLTP models in that the type II/III helium pulse extinguishes hydrogen burning drastically at the peak of the

pulse and they do not become as hydrogen deficient as VLTP models, which can reach $X \approx 0.01$ almost immediately following the very late flash. (see Lawlor & MacDonald, 2003 and 2006 for example).

In table 5 we present five timescales for key time periods, during two of which identifying an LTP might be most likely. The first period, $\Delta t_1$ is the time from a temperature $Log(T_{eff}) = 3.7$ to a time when $Log(T_{eff}) = 4.4$. The significance of this point is that this is approximately when the temperature first becomes high enough to ionize the surrounding nebula. The next period $\Delta t_2$ represents the time between $Log(T_{eff}) = 4.4$ and the models' maximum temperatures. The maximum temperature for types IIIc, IIIb, and IIIa is essentially the same as when the model reaches peak helium burning luminosity during the LTP. Once the LTP reaches a maximum in all three temperature ranges, the star rapidly declines in luminosity over 10 – 150 years while consuming hydrogen by convective mixing to some degree. The time between maximum temperature and minimum luminosity is given in table 5 and figure 4 as $\Delta t_3$. It is during this time an LTP CSPN might be most conspicuous due to the very short timescale, but also easy to overlook for the same reason. Following the rapid overall luminosity decline, a delayed reaction to the outburst causes an increase in luminosity and a decrease in temperature over timescales of only decades. Models then cool back to the AGB over several hundred years. We include these two timescales in table 5 as $\Delta t_4$ and $\Delta t_5$. The former is the time from minimum luminosity back to $Log(T_{eff}) = 4.4$ and the latter is the time from that point to minimum temperature, back to the AGB. We show a sample HR diagram with all five time periods labelled in figure 4. Last, we also show in table 5 the second crossing time as $\Delta t_6$. These are again taken from $Log(T_{eff}) = 3.7$. In all cases these times are approximately 500 to a few thousands years longer than from $Log(T_{eff}) = 4.0$ (which may be useful for comparisons). The ranges are primarily affected by stellar mass and metallicity, with larger mass and lower metals evolving the quickest. All the longest crossings are in the 0.9 M$_\odot$, Z = 0.0015 series while the shortest are for the 2.0 M$_\odot$ series.

For comparison, we also show timescales in table 5 for type V, AFTP models and type II/III thermal pulses. All timescales for type V are very short because these are closely correlated to the helium thermal pulse while still on the AGB, which itself is in general short lived. Type V models spend a very short time at temperatures higher than 25,000 K ($Log(T_{eff}) \approx 4.4$), and thus $\Delta t_1$ and $\Delta t_2$ are also quite short. Though photoionization rates can be short (Garcia et al., 2013), it may be that type V models do not remain hot long enough for the surrounding material to become photoionized. It appears in table 5 that type II/III take a longer time to evolve from maximum temperature to minimum luminosity compared with other types, but this is simply due to how late the thermal pulse occurs, after reaching maximum temperature. The timescale between the Type II/III thermal pulse peak and minimum luminosity is very short: only 5 to 7 years.

**Table 5** Range of timescales between critical evolutionary points for each model type. We include timescales for AFTP and type II/III models for comparison.

| Type | $\Delta t_1$ (yr) | $\Delta t_2$ (yr) | $\Delta t_3$ (yr) | $\Delta t_4$ (yr) | $\Delta t_5$ (yr) | $\Delta t_6$ (yr) |
|---|---|---|---|---|---|---|
| V (AFTP) | 11.4 – 650 | 8.7 – 27.5 | 6.9 – 20 | 21 – 43 | 65 – 368 | $1.2 \cdot 10^4 – 9.0 \cdot 10^4$ |
| IIIc | $4.7 \cdot 10^3 – 6.3 \cdot 10^3$ | $4.3 \cdot 10^2 – 5.7 \cdot 10^3$ | 7.4 – 43 | 46 – 114 | 141 – 254 | $2.7 \cdot 10^4 – 8.5 \cdot 10^4$ |
| IIIb | $3.6 \cdot 10^3 – 1.2 \cdot 10^4$ | $2.1 \cdot 10^3 – 1.0 \cdot 10^5$ | 18 – 150 | 52 – 490 | 181 – 375 | $2.1 \cdot 10^4 – 7.4 \cdot 10^4$ |
| IIIa | $3.1 \cdot 10^3 – 4.6 \cdot 10^3$ | $4.5 \cdot 10^3 – 1.2 \cdot 10^4$ | 12 – 72 | 82 – 193 | 116 – 239 | $7.0 \cdot 10^3 – 6.3 \cdot 10^4$ |
| II/III | $1.2 \cdot 10^3 – 2.8 \cdot 10^3$ | $2.4 \cdot 10^3 – 7.7 \cdot 10^3$ | *$2.4 \cdot 10^2 – 5.6 \cdot 10^3$ | 67 – 257 | 60 – 69 | $1.5 \cdot 10^4 – 7.7 \cdot 10^4$ |

* Although the timescale from $T_{max}$ to $L_{min}$ is long for type II/III as defined, the time range from the LTP peak to $L_{min}$ is only 5 – 7 years. The LTP peak occurs sometime after maximum temperature.

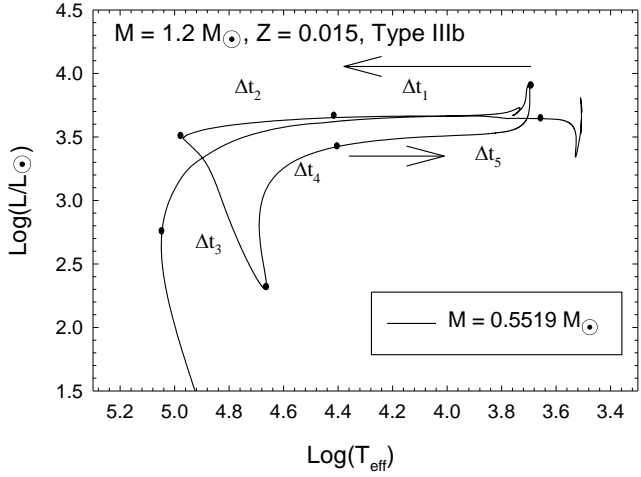

**Figure 4** A sample HR Diagram for a type IIIb LTP showing approximate timescales presented in table 5 and described in this section. Filled circles mark the boundaries of each labelled timescale. The timescale for the final (second) crossing time is also presented in table 5 as $\Delta t_6$ not labelled here for clarity.

Timescales in table 5 do not show the entire picture. The rate at which models evolve blueward across the HR diagram (the first time) can vary with temperature depending on the type of LTP. Type IIIc models initially evolve slowly from $Log(T_{eff}) = 3.6$ to $4.0$ in comparison to the evolution during higher temperatures. In this earlier period, the model evolves across the HR-diagram at a rate of 0.3 – 1.6 K/yr, as compared to 20 – 80 K/yr leading up to the helium flash peak. The evolution accelerates until its peak temperature, where the LTP also peaks. By the time it reaches 30 - 40,000 K ($Log(T_{eff}) = 4.5$ to $4.6$), the model increases in temperature at a rate of 50 K/yr. During this time, helium burning remains steady on the order of $10^3$ $L_\odot$, but gradually increasing. Still, it takes the model 120 years to heat from 40 - 50,000 K ($Log(T_{eff}) = 4.7$). Following this, the model's helium burning luminosity increases from $10^4$ $L_\odot$ to a peak around $10^6 – 10^7$ $L_\odot$ in 30 years. It subsequently declines rapidly in overall stellar luminosity over the next 7.4 – 43 years (see table 5).

The picture is similar for hotter LTP types IIIb and IIIa except their rates of evolution peak around 35,000 K ($Log(T_{eff}) = 4.55$) before slowing down by 50% near maximum temperature. But each LTP erupts in a similar way as type IIIc. Helium burning for both erupts from $10^4$ L$_\odot$ to the peak around $10^6 - 10^7$ L$_\odot$ over 30 years, beginning shortly before it reaches the LTP peak temperature shown in table 3. Overall, the slowest evolving model is our type IIIa, which also leaves the AGB with lowest helium burning luminosity. In summary, all three LTP models evolve to their peak LTP on timescales of thousands of years. They remain above Log($T_{eff}$) = 4.4 for several hundred to tens of thousands of years depending on mass. Type IIIc evolves at the highest rate, reaching 86 kK/yr right before the LTP peak. The rate for type IIIb increases until Log($T_{eff}$) = 4.6, then slows until reaching the LTP peak. Type IIIa's evolution rate remains nearly constant until it begins to speed up between Log($T_{eff}$) = 4.6 and 4.7. After, it slows to a temperature rate of increase of only 9 K/yr before the LTP rapidly develops. All three evolve very rapidly once the helium burning luminosity is above $10^4$ L$_\odot$, in about 30 years, and all drop in overall stellar luminosity between 7.4 and 150 years following the LTP peak.

## 4 BRIEF NOTES ON LTP CANDIDATES

Real LTP stars have been elusive and there are only a handful observations of possible LTP stars. In this section we briefly discuss some cases, all of which should continue to be observed and fully analysed individually.

*V839 Ara, the central star of the Stingray Nebula:*

There have been few late thermal pulse objects observed. Arguably the most likely case is V839 Ara (SAO244567). A summary and description of this star can be found in Lawlor (2021), Schaefer et al. (2021), Reindl et al. (2017, 2014), Schafer and Edwards (2015), Parthasarathy (2000), Parthasarathy et al. (1993). It was first proposed by Reindl et al. 2014 to be a likely LTP, but more models were needed to match its evolution exactly. Lawlor (2021) identified it as a type IIIc LTP (named so here), but recently, Peña et al. (2022) raised the issue that V839 Ara evolved from a B1 type post-AGB supergiant into a CSPN in the extremely short timescale of 20 years. Meanwhile, in nearly all LTP models in this work pertinent to V839 Ara, timescales for a comparable temperature increase are on the order of 100 years. This is not necessarily a problem since Miller Bertolami et al. (2016) present models that evolve more quickly than others, which they attribute primarily to convective boundary mixing used in their models. The timescales for initial heating are matched well by their models as presented in Reindl et al. (2017). We encourage and challenge observers to continue to monitor this exciting star.

*Fg Sge:*

Lawlor & MacDonald (2003) initially identified FG Sge as a VLTP where it neatly fit into a sequence beginning with the rapid redward evolution of V4334 Sgr (Sakurai's Object) and the blue then red movement of V605 Aql, cooling to the AGB. They suggested that FG Sge was completing its second trip to the AGB (ahead of V605 Aql). FG Sge was identified as a new planetary nebula in 1961 (Henize, 1961) and was identified as a nova increasing in brightness by more than four magnitudes (Richter, G., 1960). A compilation of over 100 years of observations

(van Genderen & Gautschy, 1995) of temperature and luminosity revealed a timescale consistent with cooling and brightening in VLTP models on the second approach to the AGB. At the time (2003) it was reported that FG Sge was hydrogen deficient, a pronounced effect for VLTP models. However, later Jeffery and Schönberner (2006) re-evaluated observations of FG Sge from five sources back to Herbig & Boyarchuk (1968) and Langer et al. (1974) and made a convincing argument that FG Sge was not hydrogen deficient until a point approaching minimum temperature. The second approach to the AGB for VLTP models looks qualitatively similar as the first approach during a LTP, except a VLTP is already hydrogen deficient by that time. Jeffery and Schönberner note, *"Therefore, the dilemma posed by FG Sge is that, according to the associated planetary nebula, it must have experienced a LTP. However, LTP models predict that surface hydrogen should not be depleted until a deep convection zone is established, and then only ∼ $10^4$ y after the pulse occurs. In the case of FG Sge, this hydrogen-depletion has occurred before the star reaches its minimum temperature, but substantially after the time at which it is predicted to occur following a VLTP."* Models presented in this work do not solve this problem, but one type is a more likely match than others. Only a few of our LTP models become more than a few percent deficient in hydrogen at all. Our most massive Type II/III models (2.0 $M_\odot$, Z = 0.030) that results in a remnant of M = 0.6273 $M_\odot$ becomes modestly hydrogen deficient (only 2%) just before minimum temperature. The model becomes further deficient but following $T_{min}$. Most others are reduced in hydrogen mass fraction by less than 1% - 2%. Though this does present a problem, it may be possible to solve by adjusting convective mixing efficiency during the approach to the AGB. A unique feature for the first most massive model is that it heats across the HR diagram rapidly following $T_{min}$, heating from $Log(T_{eff})$ = 3.77 to 4.4 in 200 years. Thus, this can be tested for FG Sge. If it heats rapidly, it may indicate it is a more massive type II/III helium flash object. Another possible solution is that it may well be a type II/III helium flash, but we have not yet found the right combination of initial mass and metallicity. If type II/III behavior turns out to be particular to more massive model stars, it may be FG Sge's progenitor was more massive than our models presented here, with a metallicity lower than 0.03. We plan to investigate this solution in future works.

*HD 167362 – The central star of planetary nebula SwSt 1:*

Hajduk et al., (2020) pose the question of whether HD 167362 could be a LTP in a massive post-AGB star. Though we do not include models massive enough to make a definitive determination it is more likely that this star resulted from a VLTP rather than a LTP. De Marco et al. (2001) reported that it was enriched in carbon and is hydrogen deficient. We make the VLTP argument based primarily on this, as our LTP models do not meet this requirement. What is unusual about this star is that Hajduk et al. report that the temperature decline stopped sometime between 1997 and 2015, reaching a minimum of only about $Log(T_{eff})$ = 4.6. This does not preclude it from being a VLTP entirely, as the size of AFTP, LTP and VLTP loops can vary in size depending on how much helium remains at the end of AGB. Pursuing a dense grid of models for VLTP's may be a worthwhile test.

*The central star of planetary nebula IC 49997:*

The central star of the very young planetary nebula IC 4997 had been referred to as a weak emission line star (wels) or [WC10] (Miranda et al., 2022; Hajduk et al., 2014). The classification of *wels* was first introduced by Tylenda et al. (1993). However, recently Weidmann et al. (2015, 2020) discouraged the use of wels as a classification at all, noting that in their 2015 work, most stars classified as wels were not hydrogen deficient at all. In the case of IC 4997, Weidmann (2020) reported that the presence of photospheric H cannot be decided. On the other hand, they note that the emission of He II at 4686Å is wider than nebular lines, and so may be of stellar origin. Werner & Herwig (2006) also noted that earlier Mendez (1991) and Fogel et al. (2003) reported that some wels are clearly 'usual' hydrogen rich post AGB stars. Recently, Danehkar & Parthasurathy (2022) referred to the central star as hydrogen deficient. Because the nebula is so small and dense, the nature of the central star is in question. Danehkar & Parthasurathy present the central star's temperature and luminosity between 1978 and 1991 obtained from spectral analysis and show that the star's luminosity has dropped from about $Log(L/L_\odot) = 3.35$ to 3.07, while remaining at roughly $Log(T_{eff}) = 4.7$. In the same work they compare Blocker's 1995 models, one of which resembles our type IIIa, and the second model appears to be possibly numerically peculiar. It is tempting to suggest that this star may be a type IIIc LTP based on the decline in luminosity near $Log(T_{eff}) = 4.7$ and on how young the nebula is. Both fit well with this scenario. However, there are notable difficulties. If it is indeed a [WC10], that would point to a hydrogen deficient star. Danehkar & Parthasurathy do not present luminosity following 1991, but they do present the effective temperature based on the OIII flux (λ 4959). By their analysis, the temperature has increased since 1991 to $Log(T_{eff}) \sim 4.86$ in 2019. Our type IIIc models do briefly increase in temperature, but not to quite so high a temperature ($Log(T_{eff} \sim 4.75)$. It is possible that we just have not identified the right combination of mass and metallicity for this star – perhaps a higher mass LTP IIIc model would reach a higher temperature. Continued monitoring of the temperature would be very useful in determining if an LTP can explain this star's evolution.

## 5. SUMMARY AND CONCLUSIONS

In this paper, we present five new series of late thermal pulse models. For comparisons in our grids here, we designate LTP outbursts in three ranges of temperature centred around $Log(T_{eff}) =$ 4.72, 4.85 and 4.98. We also include for comparison, models that undergo an AFTP, type V AGB departure, and models that experience a helium flash while just entering the WD cooling track but do not entirely resemble an LTP nor a VLTP (Type II/III). One potential advantage of recognizing these types is use in the identification of LTP candidates, which have to date been very sparse, or unconfirmed. We present a simple relationship between helium pulse type and helium mass remaining at the end of the AGB. When helium mass is between 2% and 4% of the stellar mass at the end of the AGB, models in our limited range of mass and metallicity experience helium flash of some variety. There is a general trend, with exceptions, that models with higher helium mass will experience a LTP at a cooler temperature. The reason this is not entirely monotonic for each model series is that it depends also on where in the thermal pulse cycle the model is when hydrogen mass drops below a critical value and the model begins to contract. By changing mass loss near the peak of the final AGB pulse, we are also adjusting where in the cycle departure occurs. For our smaller model grids, like the 1.2 $M_\odot$ and 0.9 $M_\odot$ with $Z = 0.0015$, helium mass does decrease monotonically with departure type. We also show that there is a monotonic decrease in the ratio of hydrogen mass to helium mass at the LTP peak, however this is an expected effect because hydrogen mass decreases with increasing temperature during AGB departure by mass loss. Models with the lowest helium mass at the end of the AGB evolve to become either a DA WD (type I), a VLTP (Type II) or a type II/III model, with helium mass increasing in that order. Models with the highest helium mass near the end of the AGB experience a type V AFTP or an additional thermal pulse.

      LTP's may be common during evolution between the AGB and PNe phases but finding real examples of them is problematic. Our LTP models do not create a surface that is hydrogen deficient. Only in a few cases is there a modest depletion of hydrogen following the return to the AGB. Type V helium flashes occur while the star's envelope is still attached and extended, and so there is little mixing due to the persisting entropy barrier. In all cases, no LTP becomes as hydrogen deficient as in a VLTP. The more modest hydrogen consumption bolsters the argument of Althaus et al. (2005) that DA WD's with thin hydrogen atmospheres can be explained by late thermal pulses.

      Unlike a VLTP, LTP models do not experience as dramatic a return to the AGB from the WD cooling track, increasing in luminosity conspicuously over many orders of magnitude. In the absence of a sudden brightening or an extreme composition change as for a VLTP, LTP objects present a challenge to identify. To make matters worse, LTP's take thousands of years to reach peak helium burning luminosity spending a significant time hot enough to ionize a surrounding nebula. It is not inconceivable that any planetary nebula could erupt without much warning. Our temperature ranges may hint at stellar properties such as mass or metallicity because the highest mass models spend the most time on the hot side of the HR diagram during crossing. Immediately after the helium flash reaches its peak value models rapidly decline in brightness over 10 to 150 years. In this period of decades, the evolution could be overlooked. V839 Ara, one of the only observed LTP objects was first identified as the youngest planetary nebula before its rapid decline in brightness over about 25 years. Such a discovery may require consistent monitoring over decades, or perhaps historic data mining and comparisons. Following this tiny window for near real-time visual detection, the luminosity increases over decades and finally it begins to cool back to the AGB over 200 – 900 years, slow enough that they would still be difficult to detect – particularly given that PNe have only been discovered and observed for fewer than 400 years. It is additionally possible during this return to the AGB that they may become enshrouded in dust, as has been the case for FG Sge for example (Arkhipova 1994; Woodward et al. 1993; Gonzalez et al., 1998). Thus, FG Sge type stars may also be where to

look for candidates. It does appear possible to distinguish LTP and VLTP stars at this point based on surface abundance, difficult though it may be.

**ACKNOWLEDGEMENTS**

The author thanks M. Parthasurathy for bringing new work on V839 Ara to my attention and for useful discussions about the star. Also, many thanks to an anonymous reviewer whose review and suggestions greatly improved this manuscript in numerous ways. Much of this work was made possible by Penn State Brandywine's Priscilla Clement Award for Research and The Jane Cooper Memorial Fellowship.

**DATA AVAILABILITY**

No new data were generated or analysed in support of this research.